\documentclass[aps,prc,preprint,showpacs,showkeys,byrevtex]{revtex4}
\usepackage{graphicx}
\usepackage{latexsym}
\usepackage{dcolumn}
\usepackage{bm,amsfonts,hyperref}

\newcommand{\expup}[1]{e^{#1}}

\newcommand{\Fslash}[1]{\ensuremath{\makebox[0mm][l]{\hspace{.1em}/}{#1}}}
\newcommand{\EG}{{\textrm{e.g.}}}
\newcommand{\IE}{{\textrm{i.e.}}}

\date{\today}

\begin{document}
\title{Comparison of the heavy-fermion and Foldy-Wouthuysen formalisms at
third order}

\author{A. G{\aa}rdestig}\email{anders@physics.sc.edu}
\author{K. Kubodera}\email{kubodera@sc.edu}
\author{F. Myhrer}\email{myhrer@physics.sc.edu}
\affiliation{Department of Physics and Astronomy, 
University of South Carolina, Columbia, SC 29208}

\begin{abstract}
We compare two non-relativistic (NR) reduction schemes (heavy-fermion and 
Foldy-Wouthuysen) that are used to derive low-energy effective-field-theory
Lagrangians.
We give the explicit transformation between the two types of fields to 
$\mathcal{O}(1/m^2)$, derived from a quite general, relativistic Lagrangian.
Beyond leading order the NR reductions always involve the smaller components 
of the Dirac spinors that are to be integrated out to formulate the NR theory.
Even so, the transformation between the NR Lagrangians can be carried out 
explicitly to $\mathcal{O}(1/m^2)$ using a field renormalization, as long as 
the lower components of the Lagrangian are known.
The fixed coefficient corrections to some low-energy constants at 
$\mathcal{O}(1/m^2)$ will depend on the particular scheme chosen, but will 
match after the field renormalization.
\end{abstract}


\maketitle
\section{Introduction}
When considering low-energy phenomena, it often simplifies the calculations to
start from a non-relativistic Lagrangian.
Several useful and well-known examples can be found.
These include the low-energy few-body nuclear systems for which chiral 
perturbation theory ($\chi$PT) has been developed, the electromagnetic 
properties of composite particles (moments of the proton and neutron), atomic
physics [non-relativistic quantum electrodynamics (NRQED)] and heavy-quark 
effective field theory (HQEFT).
Since the original theory, here assumed to be an effective field theory (EFT) 
in the Weinberg sense~\cite{Weinberg}, is generally given by a relativistic
Lagrangian, one faces the problem of performing a non-relativistic (NR) 
reduction.

Several solutions have been suggested over the years.
The simplest, which we will label the direct Pauli reduction scheme or the
$1/m$ expansion, expands the Dirac matrix element in terms of the 
two-dimensional Pauli-spinors that appear in the usual representation of 
the four-spinors.

HQEFT was originally formulated in the non-relativistic limit, 
utilizing the constraints given by the symmetries of quantum chromodynamics 
(QCD) in the static limit~\cite{Georgi}.
Later HQEFT was elegantly derived from the relativistic QCD Lagrangian
using a path-integral method~\cite{Mannel}; this we will call the 
heavy-fermion (HF) approach (for a recent review of HQEFT, see 
\cite{Brambilla1}).
These ideas were carried over to the descriptions of heavy 
baryon chiral perturbation theory (HB$\chi$PT)~\cite{JM,BKKM,ulfreview}.

An alternative NR reduction method would be the classic 
Foldy-Wouthuysen (FW) transformation~\cite{FW,BD}, which was designed to 
provide the NR limit of quantum electrodynamics (QED).
It has also been used to derive HQEFT~\cite{KT}, but this work was initially
criticized by Ref.~\cite{Mannel} for not proving HQEFT to be an EFT (see, 
however, later work~\cite{Mannel2,Korner}).
According to Ref.~\cite{Mannel}, HF and FW give identical 
results for HQEFT, although as far as we are able to ascertain, this was 
shown only to $\mathcal{O}(1/m)$, where $m$ is the (heavy) fermion mass.
Later, these two methods were indeed shown to be equivalent, after a wave 
function renormalization and a unitary transformation \cite{DasMathur,Das}
(see also Refs.~\cite{Holstein,Brambilla2}).
However, these references considered only renormalizable theories (QED and QCD)
and \cite{Das} only the explicit example of a constant electric field (in QED).
Neither considered the low-energy constants (LECs), which parametrize the 
(usually unknown) short-distance physics in an EFT like, \EG, $\chi$PT.

It has already been shown that, when the relativistic theory is fully 
known, the direct Pauli reduction scheme ($1/m$ expansion) differs from FW and 
only the latter gives $S$-matrices that are in agreement with the relativistic 
ones~\cite{FPS}.
This provided part of the inspiration for the present investigation.
Another motivation was the claims in the literature that FW and HF are 
equivalent for HQEFT~\cite{Mannel,KT} as well as for HB$\chi$PT~\cite{APKM}.

In both HF and FW the contributions involving the lower Dirac spinor components
are eliminated, and this can be formalized by integrating out these ``small'' 
components from the path integral.
For momenta and field energies smaller than $m$, the surviving terms can be
expanded in a perturbative series.
The difference between the two methods lies in the way the separation of lower
and upper components is made, resulting in different NR fields.
Note that this `integrating-out' procedure is different from what is done in, 
\EG, electroweak theory, where the heavy vector bosons are integrated out 
completely, at lowest order giving the Fermi weak interaction model. 
Here we rather want to keep the heavy (fermion) degrees of freedom (when they 
are protected by a symmetry, like baryon number conservation), but eliminate 
the explicit antifermion contributions (lower components), since they can not 
be excited in the non-relativistic limit.

In this paper we will study the HF and FW representations in detail, starting 
from a quite general, relativistic, Lagrangian, which is assumed to be known.
The two methods appear to disagree at higher orders in $1/m$,
\IE, the so-called ``fixed-coefficient'' terms ($1/m$ corrections) differ 
between the Lagrangians.
We will give explicit expressions for the transformation necessary
to get from HF to FW to $\mathcal{O}(1/m^2)$, showing that the corresponding 
fields each contain pieces of the small/lower components of the other, already 
at $\mathcal{O}(1/m)$.
Even so, the functional form of the corresponding Lagrangians will differ first
at $\mathcal{O}(1/m^2)$, which explains the equivalence found between the
$\mathcal{O}(1/m)$ results of Refs.~\cite{Mannel} and \cite{KT}.
It is possible to show the equivalence of the two approaches~\cite{Das}
(at least for a constant electric field in NRQED), by using field 
renormalization and unitary transformations.
We want to treat a more general Lagrangian than what was done in~\cite{Das} 
and will show that at higher orders there are differences in the 
fixed-coefficient ($1/m$) corrections to the LECs in, \EG, $\chi$PT.
(In this paper we will only be concerned with LECs introduced already in the 
relativistic Lagrangian, as done in, \EG, \cite{ulfreview,Fettes1,Fettes2}.)
Thus, one has to be careful to use correctly renormalized fields when
comparing LECs at this order.
At higher order also unitary transformations need to be applied.

This paper is organized as follows:
In Sec.~\ref{sec:NRR} we give a brief review of the essential features of the
HF and FW schemes, followed by a detailed comparison and derivation of the
transformation between them.
The paper is concluded by a discussion in Sec.~\ref{sec:concl}.
Some mathematical details are collected in an Appendix.

\section{Non-relativistic reductions}
\label{sec:NRR}
We will assume as the starting point the relativistic Lagrangian
\begin{equation}
  \mathcal{L} = \bar\psi\left(i\Fslash{\mathcal{D}}-m+\gamma_0G\right)\psi,
\label{eq:Lorig}
\end{equation}
where the covariant derivative $\mathcal{D}^\mu\equiv\partial^\mu+\Gamma^\mu$
and the general coupling matrix $G$ (the extra factor of $\gamma_0$ is included
for future convenience) is
\begin{eqnarray}
  G & \equiv & \left( \begin{array}{cc} A & B \\ C & D \end{array} \right),
\end{eqnarray}
where $A$--$D$ are $2\times2$ matrices.
The connection $\Gamma^\mu$ and the coupling matrix $G$ contain all
single-fermion couplings (to, \EG, pions and electroweak fields) allowed by 
relevant symmetries, \EG, chiral symmetry, Lorentz invariance, parity, flavor
and color SU(3) etc.
Throughout the paper we will work in the usual Dirac representation~\cite{BD} 
and $G$ should be interpreted in that representation. 
We further assume that $G$ is hermitian, \IE, that $A$ and $D$ are hermitian 
and $B^\dagger=C$.
No assumptions will be made regarding the commutative properties of the 
elements of $G$ and $\mathcal{D}$, so the results of this paper apply
equally well to Abelian and non-Abelian field theories.
Note that $A$--$D$ and $\mathcal{D}^\mu$ will in general contain higher order
interactions terms which should be expanded in the final NR formulas.
This expansion is separate from the $1/m$ expansion created by the NR 
reduction, but in some cases the two expansions can be combined into a single
counting scheme.
This happens for, \EG, HB$\chi$PT, where $m\sim\Lambda_\chi$ and
$\Lambda_\chi\sim1$~GeV is the chiral scale.
The generating functional for the Lagrangian~(\ref{eq:Lorig}) is
\begin{equation}
  \mathcal{Z} = \int[d\bar\psi][d\psi][dX]
  \exp\left(-i\int d^4x\mathcal{L}\right),
\end{equation}
where $X$ represents the fields implicit in $\Gamma^\mu$ and $G$.
These fields will from now on be suppressed in our notation.

\subsection{A brief overview of the heavy-fermion method}
\label{sec:HB}
If the fermion is heavy, we can for low four-momenta treat it as essentially 
static and expand around its large mass.
Following Ref.~\cite{Georgi}, this is accomplished by writing the fermion 
momentum $p^\mu$ in terms of the four-velocity $v^\mu$ and a residual fermion 
momentum $l^\mu$ as
\begin{equation}
  p^\mu = mv^\mu+l^\mu,
\end{equation}
where $l^\mu\ll m$ and $v^2=1$.
Consequently, for an on-shell fermion, $p^2=m^2$ implies that
\begin{equation}
  2mv\cdot l+l^2 = 0.
\label{eq:onshell}
\end{equation}
The fermion field is next split into large ($H$) and small ($h$) components
\begin{eqnarray}
  \psi & = & \expup{-imv\cdot x}(H+h),
\end{eqnarray}
where $\Fslash{v}H=H$ and $\Fslash{v}h=-h$, which shifts all the linear 
mass dependence to the small component.
Without loss of generality we will assume for the remainder of this article
that $v^\mu=(1,0,0,0)$, and thus $\Fslash{v}=\gamma_0$, which simplifies the
algebra and the comparison with FW.
(It is possible to keep $v^\mu$ general and also do the FW transformation in 
terms of $v^\mu$~\cite{KT,DasMathur}, but we choose a simpler approach here.)
The resulting Lagrangian is
\begin{eqnarray}
  \mathcal{L} & = & 
  \left(\begin{array}{cc} H^\dagger & h^\dagger \end{array} \right) 
  \left(\begin{array}{cc} 
    i\mathcal{D}_0+A & -i\vec\sigma\cdot\vec\mathcal{D}+B \\
    -i\vec\sigma\cdot\vec\mathcal{D}+C & 2m+i\mathcal{D}_0+D
  \end{array}\right)
  \left(\begin{array}{c} H \\ h \end{array}\right),
\end{eqnarray}
where we have chosen to work with $H^\dagger$ instead of $\bar{H}$ to ease the
comparison with the FW expressions further on.
The cross talk between upper and lower components can be eliminated by defining
a new small component field $h'$~\cite{Mannel} such that
\begin{eqnarray}
  h & = & h'-(2m+i\mathcal{D}_0+D)^{-1}(-i\vec\sigma\cdot\vec\mathcal{D}+C)H, 
  \nonumber \\
  h^\dagger & = & {h'}^\dagger-
  H^\dagger(-i\vec\sigma\cdot\vec\mathcal{D}+B)(2m+i\mathcal{D}_0+D)^{-1}.
\label{eq:sq}
\end{eqnarray}
This turns the Lagrangian into a block-diagonal form:
\begin{equation}
  \mathcal{L} = H^\dagger\left(i\mathcal{D}_0+A
  -(-i\vec\sigma\cdot\vec\mathcal{D}+B)
  \frac{1}{2m+i\mathcal{D}_0+D}(-i\vec\sigma\cdot\vec\mathcal{D}+C)\right)H
  +{h'}^\dagger(2m+i\mathcal{D}_0+D)h', \nonumber \\
\label{eq:HHhh}
\end{equation}
and `completes the square' in the path integral exponential.
Up to this point no approximations have been made, so this equation can be
regarded as a Dirac equation in two-component form.
After `integrating out' the small components, which gives a constant 
determinant~\cite{Mannel} (see the Appendix), only the $H$ fields remain 
and we have obtained a rudimentary form of the HF Lagrangian.
Note that the $H$ fields have to be renormalized to preserve the
norm~\cite{Das,Okubo}.
The antifermion is implicitly included as the `$z$-graph'--like term of 
Eq.~(\ref{eq:HHhh}).
We next expand this term, assuming that $i\mathcal{D}^\mu,G\ll 2m$.
Thus, this part of the HF Lagrangian is expanded in powers of 
$(i\mathcal{D}_0+D)/2m$, which contains only time derivatives of the fermion
field (apart from possible derivatives in $G$), sandwiched between two space 
derivatives.
The first few orders of the HF Lagrangian are then
\begin{eqnarray}
  \mathcal{L}^{(0)}_{\rm HF} & = & H^\dagger(i\mathcal{D}_0+A)H, \\
  \mathcal{L}^{(1)}_{\rm HF} & = & -\frac{1}{2m}
  H^\dagger(-i\vec\sigma\cdot\vec\mathcal{D}+B)
  (-i\vec\sigma\cdot\vec\mathcal{D}+C)H, \\
  \mathcal{L}^{(2)}_{\rm HF} & = & \frac{1}{4m^2}
  H^\dagger (-i\vec\sigma\cdot\vec\mathcal{D}+B)(i\mathcal{D}_0+D)
  (-i\vec\sigma\cdot\vec\mathcal{D}+C)H,
\label{eq:HF}
\end{eqnarray}
where the indices indicate inverse powers of $m$.
The Lagrangian in general also contains higher order interaction terms in 
$A$--$D$ and $\mathcal{D}^\mu$. 
This expansion will not be shown explicitly.
For the purposes of this paper we count 
$\mathcal{D}_0\sim l_0\sim|\mathbf{l}|$, although counting 
$\mathcal{D}_0\sim \mathbf{l}^2/m$ is also possible, depending 
on the process being studied.
Short-range interactions are parametrized as local counter terms, which
contain LECs.
The LECs are determined either by matching conditions (when the underlying,
more complete theory is well known, \EG, HQEFT and non-relativistic QED 
(NRQED)~\cite{NRQED}) or by fitting EFT calculations to data (when the 
underlying theory is not well known, \EG, HB$\chi$PT).
In some cases, models such as resonance saturation are used to evaluate or 
estimate the LECs.

The propagator for the $H$ field is given by Eq.~(\ref{eq:HHhh})
\begin{equation}
  \frac{i}{i\partial_0+\nabla^2\frac{1}{2m+i\partial_0}} = 
  \frac{i(1+\frac{i\partial_0}{2m})}{i\partial_0-\frac{\Box}{2m}},
\label{eq:Hprop}
\end{equation}
where $\Box=\partial_0^2-\nabla^2$.
The inverse of this propagator is non-linear in $i\partial_0$ beyond a certain 
order, and this propagator usually appears in expanded 
form~\cite{ulfreview,Fettes1,Fettes2}.
In order to compare this HF propagator to the FW propagator that we will give
below, we choose a situation that appears often in our own line of research, 
\IE, pion-nucleon dynamics in HB$\chi$PT.
Consider an on-shell nucleon with residual four-momentum $l$.
The HF nucleon propagator after it has emitted one pion of four-momentum $q$ 
(real or virtual, but with $q_0,|{\bf q}|\ll m$) can be written as
\begin{eqnarray}
  \frac{i\left(1+\frac{l_0-q_0}{2m}\right)}{l_0-q_0+\frac{(l-q)^2}{2m}} 
  & = & \frac{i\left(1+\frac{l_0-q_0}{2m}\right)}{-q_0-\frac{l\cdot q}{m}+
    \frac{q^2}{2m}},
\label{eq:HFprop}
\end{eqnarray}
where the relativistic on-shell condition~(\ref{eq:onshell}) has been applied.
The propagator (\ref{eq:HFprop}) can be further expanded as needed, whether in 
$1/m$ or in $\sqrt{m_\pi/m}$ as in pion production off two 
nucleons~\cite{Cohen}.
This application of the propagator is similar to the one advocated
by Hanhart and Wirzba~\cite{HW}.
After a simple re-identification of $l$ and $q$ as the electron and photon
four-momenta, Eq.~(\ref{eq:HFprop}) can also represent the NRQED electron 
propagator after emission of a soft photon.

The HF propagator [Eq.~(\ref{eq:HFprop})] is identical to an $1/m$ expansion of
the relativistic propagator, projected on the upper left 2$\times$2 submatrix:
\begin{eqnarray}
  \frac{i(m\Fslash{v}+\Fslash{\,l}-\Fslash{q}+m)}{(mv+l-q)^2-m^2} & \to &  
  i\frac{2m+l_0-q_0}{-2(m+l_0)q_0+2\mathbf{l}\cdot\mathbf{q}+q^2} 
  \nonumber \\
  & = &  i\frac{1+\frac{l_0-q_0}{2m}}
	    {-q_0-\frac{l\cdot{q}}{m}+\frac{q^2}{2m}}.
\label{eq:NRprop}
\end{eqnarray}

\subsection{Foldy-Wouthuysen transformation in a Lagrangian formulation}
\label{sec:FW}
In order to compare to the HF results we choose to derive the FW transformation
from the Lagrangian~\cite{KT,Holstein} instead of the Hamiltonian as is 
usually done.
In addition this approach makes it possible to treat the time and space 
components in a more symmetrical manner.
The time derivative is implicit in all expressions containing 
$\mathcal{D}_0$.
Our starting point is the Lagrangian~(\ref{eq:Lorig}) rewritten as
\begin{equation}
  \mathcal{L} = \psi^\dagger\left( i\mathcal{D}_0
  -i\gamma_0\vec\gamma\cdot\vec\mathcal{D}-\gamma_0m+G\right)\psi,
\end{equation}
where the field $\psi^\dagger$ is chosen instead of $\bar\psi$ since it makes 
the algebra somewhat more transparent for the present purposes.
This expression can be further separated into odd and even terms, \IE, terms 
that mix and do not mix the upper and lower components (in the Dirac 
representation)
\begin{equation}
  \mathcal{L} = \psi^\dagger(-\mathcal{E}-\mathcal{O}-\gamma_0m)\psi,
\end{equation}
where the sign convention for the even ($\mathcal{E}$) and odd ($\mathcal{O}$)
operators is in accordance with Bjorken-Drell~\cite{BD}.
The mass term can be removed from the upper components by shifting the 
zero-point of the energy, \IE, we define $\tilde\psi=\expup{-imt}\psi$, giving
\begin{equation}
  \mathcal{L} = \tilde\psi^\dagger[-\mathcal{E}-\mathcal{O}-(\gamma_0-1)m]
  \tilde\psi.
\end{equation}
The explicit expressions for the even and odd operators are
\begin{eqnarray}
  \mathcal{E} & \equiv & -\left( \begin{array}{cc} i\mathcal{D}_0+A & 0 \\
  0 & i\mathcal{D}_0+D \end{array} \right), \\
  \mathcal{O} & \equiv & -\left( 
  \begin{array}{cc} 0 & -i\vec\sigma\cdot\vec\mathcal{D}+B \\
  -i\vec\sigma\cdot\vec\mathcal{D}+C & 0 \end{array} \right).
\end{eqnarray}

The FW transformation redefines the fermion field through the consecutive
application of unitary matrices: 
\begin{eqnarray}
  \psi''' & = & \expup{iS''}\expup{iS'}\expup{iS}\tilde\psi, \label{eq:psi3} \\
  S & = & -\frac{i\gamma_0\mathcal{O}}{2m},
\label{eq:SO}
\end{eqnarray}
where $S'$ is related to $\mathcal{O}'$ in the same way as $S$ to $O$ 
[Eq.~(\ref{eq:SO})] and similarly for $S''$ and $\mathcal{O}''$.
These higher order odd operators are introduced below.
The transformations carried out to $\mathcal{O}(1/m^3)$ lead to the 
Lagrangian~\cite{BD}
\begin{eqnarray}
  \mathcal{L}''' & = & \psi'''^\dagger \left(-\mathcal{E}'-(\gamma_0-1)m
  \right)\psi''', \\
  \mathcal{E}' & = & \gamma_0\left(\frac{\mathcal{O}^2}{2m}
  -\frac{\mathcal{O}^4}{8m^3}\right)+\mathcal{E}
  -\frac{1}{8m^2}[\mathcal{O},[\mathcal{O},\mathcal{E}]], \label{eq:Ep} \\
  \mathcal{O}' & = & \frac{\gamma_0}{2m}[\mathcal{O},\mathcal{E}]
  -\frac{\mathcal{O}^3}{3m^2}, \\
  \mathcal{O}'' & = & \frac{\gamma_0}{2m}[\mathcal{O}',\mathcal{E}'],
\end{eqnarray}
where it is assumed that the time derivative in $\mathcal{E}$, 
\IE, in $\mathcal{D}_0$, acts on everything to its right.
All odd terms have been transformed away to this order and only
the even operators remain.
The explicit expressions for the higher order odd operators are
\begin{eqnarray}
  \mathcal{O}' & = & \frac{1}{2m}\left( \begin{array}{cc} 
    0 & B' \\
    -C' & 0
  \end{array}\right), \\
  \mathcal{O}'' & = & -\frac{1}{4m^2} \left( \begin{array}{cc} 
    0 & B'(i\mathcal{D}_0+D)-(i\mathcal{D}_0+A)B' \\
    C'(i\mathcal{D}_0+A)-(i\mathcal{D}_0+D)C' & 0
  \end{array}\right),
\end{eqnarray}
where we have defined
\begin{eqnarray}
  B' & = & (-i\vec\sigma\cdot\vec\mathcal{D}+B)(i\mathcal{D}_0+D)-
  (i\mathcal{D}_0+A)(-i\vec\sigma\cdot\vec\mathcal{D}+B), \\
  C' & = & (-i\vec\sigma\cdot\vec\mathcal{D}+C)(i\mathcal{D}_0+A)-
  (i\mathcal{D}_0+D)(-i\vec\sigma\cdot\vec\mathcal{D}+C).
\end{eqnarray}
The upper ($N$) and lower ($n$) FW components of $\psi'''$ 
[Eq.~(\ref{eq:psi3})] are related to the HF fields $H$ and $h$ by the 
transformation
\begin{eqnarray}
  \left(\begin{array}{c} N \\ n \end{array}\right) & = &
  \left(\begin{array}{cc} 
    1-\frac{1}{8m^2}\tilde{B}\tilde{C}+\frac{1}{8m^3}B'\tilde{C}
    & -\frac{1}{2m}\tilde{B}+\frac{1}{4m^2}B'
    +\frac{3}{16m^3}\tilde{B}\tilde{C}\tilde{B}-\frac{1}{8m^3}B'' \\
     \frac{1}{2m}\tilde{C}+\frac{1}{4m^2}C'
     -\frac{3}{16m^3}\tilde{C}\tilde{B}\tilde{C}+\frac{1}{8m^3}C''
     & 1-\frac{1}{8m^2}\tilde{C}\tilde{B}-\frac{1}{8m^3}C'\tilde{B}
  \end{array}\right)
  \left(\begin{array}{c} H \\ h \end{array}\right), \nonumber \\
\end{eqnarray}
where we use the abbreviated notation
\begin{eqnarray}
  \tilde{B} & \equiv & -i\vec\sigma\cdot\vec\mathcal{D}+B, \\
  \tilde{C} & \equiv & -i\vec\sigma\cdot\vec\mathcal{D}+C, \\
  B'' & \equiv & 
  B'(i\mathcal{D}_0+D)-(i\mathcal{D}_0+A)B', \\
  C'' & \equiv &
  C'(i\mathcal{D}_0+A)-(i\mathcal{D}_0+D)C'.
\end{eqnarray}
The Lagrangian after the three FW transformations is
\begin{eqnarray}
  \mathcal{L}^{(-1)}_{\rm FW} & = & 2mn^\dagger n, \label{eq:FW-1} \\
  \mathcal{L}^{(0)}_{\rm FW} & = & 
  N^\dagger(i\mathcal{D}_0+A)N+n^\dagger(i\mathcal{D}_0+D)n, \label{eq:FW0} \\
  \mathcal{L}^{(1)}_{\rm FW} & = & -\frac{1}{2m} N^\dagger 
  (-i\vec\sigma\cdot\vec\mathcal{D}+B)(-i\vec\sigma\cdot\vec\mathcal{D}+C)N
  \nonumber \\
  & + & \frac{1}{2m}n^\dagger(-i\vec\sigma\cdot\vec\mathcal{D}+C)
  (-i\vec\sigma\cdot\vec\mathcal{D}+B)n, \label{eq:FW1} \\
  \mathcal{L}^{(2)}_{\rm FW} & = & \frac{1}{4m^2}
  N^\dagger (-i\vec\sigma\cdot\vec\mathcal{D}+B)(i\mathcal{D}_0+D)
  (-i\vec\sigma\cdot\vec\mathcal{D}+C) N \nonumber \\
  & - & \frac{1}{8m^2} N^\dagger \{(-i\vec\sigma\cdot\vec\mathcal{D}+B)
  (-i\vec\sigma\cdot\vec\mathcal{D}+C),(i\mathcal{D}_0+A)\} N \nonumber \\
  & + & \frac{1}{4m^2}
  n^\dagger(-i\vec\sigma\cdot\vec\mathcal{D}+C)(i\mathcal{D}_0+A)
  (-i\vec\sigma\cdot\vec\mathcal{D}+B)n \nonumber \\
  & - & \frac{1}{8m^2}n^\dagger\{(-i\vec\sigma\cdot\vec\mathcal{D}+C)
  (-i\vec\sigma\cdot\vec\mathcal{D}+B),(i\mathcal{D}_0+D)\}n.
\label{eq:FW2}
\end{eqnarray}
By construction terms mixing upper and lower components start to appear only at
the next higher order, so the square is already `completed' and one can 
integrate out the $n^\dagger n$ contributions to this order.
As in the HF case, the integration over the small components will give a 
constant determinant (see the Appendix). 
When higher order terms, involving higher order LECs, are included in the 
relativistic Lagrangian, one has to take into account that $A$--$D$ and 
$\mathcal{D}^\mu$ contain higher order interaction terms, such that further
expansion of these quantities becomes necessary.
Just like HF, the FW transformation generates an expansion in $1/m$, but the 
two series are not identical.

For example, the free $N$ propagator is (see Ref.~\cite{BD})
\begin{equation}
  \frac{i}{i\partial_0-\left(\sqrt{m^2-\nabla^2}-m\right)}=
  \frac{i}{i\partial_0+\frac{\nabla^2}{2m}+\frac{\nabla^4}{8m^3}+\ldots},
\label{eq:Nprop}
\end{equation}
\IE, exactly what would naively be expected for a non-relativistic propagator.
Unlike the inverse of the HF propagator (\ref{eq:Hprop}) the inverse of 
(\ref{eq:Nprop}) is linear in $i\partial_0$, although its dependence on the
spatial components is more complicated.
Applying this FW propagator to the same situation as in the HF case 
(but with the relativistic on-shell condition now being 
$l_0=\sqrt{m^2+\mathbf{l}^2}-m=
\frac{\mathbf{l}^2}{2m}-\frac{\mathbf{l}^4}{8m}\ldots$) we get
\begin{equation}
  \frac{i}{l_0-q_0-\frac{(\mathbf{l}-\mathbf{q})^2}{2m}+
    \frac{(\mathbf{l}-\mathbf{q})^4}{8m^3}+\ldots} = 
  \frac{i}{-q_0-\frac{\mathbf{q}^2-2\mathbf{l}\cdot\mathbf{q}}{2m}+
    \frac{(\mathbf{l}-\mathbf{q})^4-\mathbf{l}^4}{8m^3}+\ldots},
\end{equation}
which differs from the HF propagator [Eq.~(\ref{eq:HFprop})].
Although the HF and FW propagators go on-shell for the same value of $l^\mu$ 
for any given order, in general they do differ off-shell beyond leading order.
Also the on-shell residues are different, as discussed already in \cite{Das}.
It was pointed out in Ref.~\cite{Das}, however, that the difference between 
these propagators can be compensated for by a field redefinition.

\subsection{Comparison of HF and FW to third order}
\label{sec:comp}
In contrast to HF, FW is an expansion in terms of both space and time 
derivatives of the fermion, \IE, the FW transformation is not restricted to the
time derivatives only.
The two schemes also differ in that the integrated-out pieces of FW have 
contributions at all orders, not just the first few as in HF---compare 
Eq.~(\ref{eq:HHhh}) to Eqs.~(\ref{eq:FW-1})--(\ref{eq:FW2}).
Comparison with Eq.~(\ref{eq:HF}) shows that the difference in the Lagrangian
between these two expansion choices is given (to this order) by the second 
line of Eq.~(\ref{eq:FW2}), \IE, differences start to appear at 
$\mathcal{O}(1/m^2)$.
Both result in expansions in $1/m$, where $m$ indicates the size of the
cutoff of the theory.
Note that the fields $H$ and $N$ are not identical (and neither are
$h'$ and $n$), except at the lowest order, but are related through a rather 
involved transformation, see below.
In contrast to the transformation in \cite{Das}, which was carried out to 
$\mathcal{O}(1/m^5)$ for a constant electric field, we have kept the Lagrangian
general (not necessarily renormalizable, \EG, HB$\chi$PT), but restricted the 
expansion to $\mathcal{O}(1/m^2)$.

In order to transform from the HF to FW formalisms we need to `complete the
square' for the $H$ and $h$ fields according to Eq.~(\ref{eq:sq}). 
This boils down to an additional transformation
\begin{equation}
  \left(\begin{array}{c} H \\ h \end{array}\right)
  = \left( \begin{array}{cc} 1 & 0  \\
  -\frac{1}{2m+i\mathcal{D}_0+D}(-i\vec\sigma\cdot\vec\mathcal{D}+C) 
  & 1 \end{array} \right)
  \left(\begin{array}{c} H \\ h' \end{array}\right).
\label{eq:blockdiag}
\end{equation}
We then arrive at the field transformation
\begin{equation}
  \psi^{(i)} = \left( \begin{array}{c} N^{(i)} \\ n^{(i)} \end{array}\right)
  =M^{(i)}\left( \begin{array}{c} H \\ h' \end{array}\right),
\label{eq:Mdef}
\end{equation}
where
\begin{eqnarray}
  M^{(1)} & = & \left( \begin{array}{cc}
    1 & -\frac{1}{2m}\tilde{B} \\
    0 & 1 
  \end{array} \right), \label{eq:M1} \\
  M^{(2)} & = & \left( \begin{array}{cc} 
    1+\frac{1}{8m^2}\tilde{B}\tilde{C} & 
    -\frac{1}{2m}\tilde{B}+
    \frac{1}{4m^2}B' \\
    \frac{1}{4m^2}\tilde{C}(i\mathcal{D}_0+A)
    &
    1-\frac{1}{8m^2}\tilde{C}\tilde{B}
  \end{array} \right), \label{eq:M2} \\
  M^{(3)}-M^{(2)} & = & \left( \begin{array}{cc} 
    -\frac{1}{8m^3}\tilde{B}(i\mathcal{D}_0+D)\tilde{C} & 
    \frac{3}{16m^3}\tilde{B}\tilde{C}\tilde{B}-\frac{1}{8m^3}B'' \\
    -\frac{1}{8m^3}\tilde{C}\tilde{B}\tilde{C}
    +\frac{1}{8m^3}[C'-(i\mathcal{D}_0+D)\tilde{C}](i\mathcal{D}_0+A)
    &
    -\frac{1}{8m^3}C'\tilde{B}
  \end{array} \right). \nonumber \label{eq:M3} \\
\end{eqnarray}
We have checked that this transformation indeed transforms the Lagrangian from
the FW to the HF representation [to $\mathcal{O}(1/m^2)$].
This is \emph{not} a unitary transformation, because of the 
block-diagonalizing done in Eq.~(\ref{eq:blockdiag}).
It is equivalent to the wave function renormalization and unitary
field redefinitions in Refs.~\cite{Das,Okubo}.
In contrast to Ref.~\cite{Das}, however, we give the general transformation to 
$\mathcal{O}(1/m^2)$, \IE, it applies equally well to non-Abelian and 
non-renormalizable theories. 
Obviously, the FW spinor $N$ contains parts of the (to be) integrated-out
HF spinor $h'$.
This means that transforming the upper/large components from one to the 
other mixes in lower components of the other representation.
Thus, the transformation can be carried out explicitly only in cases where 
the lower component Lagrangian is known (up to the required order).

The general relation between the HF and FW Lagrangians to a given order is
\begin{equation}
  H^\dagger U_{HH}H+{h'}^\dagger L_{h'h'}h'=N^\dagger U_{NN}N
  +n^\dagger L_{nn}n.
\end{equation}
Assuming a general form for the $i$th order matrix
$M^{(i)}=\left(\begin{array}{cc} a^{(i)} & b^{(i)} \\ c^{(i)} & d^{(i)} 
\end{array}\right)$ that relates the FW and HF fields [Eq.~(\ref{eq:Mdef})] 
we can write [hereafter suppressing the index $(i)$]
\begin{eqnarray}
  \left( \begin{array}{cc} 
    U_{HH} & 0 \\ 0 & L_{h'h'}
  \end{array} \right) & = & 
  \left( \begin{array}{cc}
    a^\dagger & c^\dagger \\ b^\dagger & d^\dagger
  \end{array} \right)
  \left( \begin{array}{cc} 
    U_{NN} & 0 \\ 0 & L_{nn}
  \end{array} \right)
  \left(\begin{array}{cc} a & b \\ c & d \end{array}\right).
\end{eqnarray}
This puts constraints on the Lagrangian operators $U_{XX}$ and $L_{xx}$:
\begin{eqnarray}
  U_{HH} & = & a^\dagger U_{NN}a+c^\dagger L_{nn}c, \\
  L_{h'h'} & = & b^\dagger U_{NN}b+d^\dagger L_{nn}d, \\
  0 & = & a^\dagger U_{NN}b+c^\dagger L_{nn}d, \label{eq:abcd} \\
  0 & = & b^\dagger U_{NN}a+d^\dagger L_{nn}c. \label{eq:badc}
\end{eqnarray}
By taking advantage of the block-diagonal conditions [(\ref{eq:abcd}) and 
(\ref{eq:badc})], the lower component $L_{nn}$ can be eliminated from the 
expression for $U_{HH}$ and we arrive at a relation between the upper 
left blocks of the two Lagrangians:
\begin{equation}
    U_{HH} = a^\dagger U_{NN}(a-bd^{-1}c)=
    [a^\dagger-c^\dagger(d^\dagger)^{-1}b^\dagger]U_{NN}a.
\end{equation}
This gives a relation between the operator forms of the Lagrangians that
remain after integrating out the respective lower components.
To $\mathcal{O}(1/m^2)$ this expression is simplified to
\begin{equation} 
  U_{HH} = a^\dagger U_{NN}a.
\label{eq:Urel}
\end{equation}
The explicit expression as provided by Eq.~(\ref{eq:M2}) is
\begin{equation}
  U_{HH} = \left(1+\frac{1}{8m^2}\tilde{B}\tilde{C}\right)
  U_{NN}\left(1+\frac{1}{8m^2}\tilde{B}\tilde{C}\right),
\end{equation}
which is the general form of the wave function renormalization to this order.
It reproduces, \EG, the special cases considered in \cite{Das}.
Incidentally, this expression explains why the $\mathcal{O}(1/m)$ expressions
look equivalent---the transformation between the HF and FW Lagrangians
is $\mathcal{O}(1/m^2)$.

The $1/m$ corrections to contact terms can be different for the two methods.
This can be seen in, \EG, the $\pi N$ Lagrangian that has been developed to 
$\mathcal{O}(1/m^3)$~\cite{EM,Fettes1} and higher~\cite{Fettes2} in HB$\chi$PT,
using the HF method.
We find that the $g_A^3/m^2$ corrections to the LECs $\hat{d}_{12}$ and 
$\hat{d}_{13}$ in \cite{Fettes1,Fettes2} are different in the HF and FW
formalisms.
Thus, if these LECs are calculated, \EG, by resonance saturation, or extracted 
from data using the relativistic formulation, it would seem
like the corresponding $1/m$ corrections would matter.
This difference could be absorbed by a field renormalization and 
unitary transformation, see Refs~\cite{Das,Okubo}, provided the lower
components of the relativistic Lagrangian are known.
Therefore, the equivalence of the HF and FW approaches can be established quite
generally and in particular regarding the values of LECs at higher orders.
Since the FW result reproduces the relativistic $S$-matrix~\cite{FPS}, it is 
clear that so will the corresponding HF result.

\section{Conclusions}
\label{sec:concl}
We have carried out the heavy-fermion (HF) and Foldy-Wouthuysen (FW) 
non-relativistic reduction schemes to $\mathcal{O}(1/m^2)$, starting from a 
quite general, relativistic, Lagrangian.
At this order explicit differences start to appear in the expressions of
the respective non-relativistic Lagrangians.
Our investigation treats a more general Lagrangian than the one in \cite{Das}
and is especially concerned with the LECs (introduced in the relativistic 
Lagrangian) as they appear in the NR theory, \EG, $\chi$PT.

We have derived the field transformation between the two descriptions 
to $\mathcal{O}(1/m^2)$ and have shown how the $1/m$ corrections to certain
LECs differ between the two approaches at this order.
Although the Lagrangians differ at $\mathcal{O}(1/m^2)$, they can be related 
through a field redefinition, which we give in a more general form than 
the one in Ref.~\cite{Das}.
Thus, when using LECs of higher order, one must ascertain that the fields
are appropriately redefined.

We have also shown the equivalence between the NR propagators derived in HF 
and from an $1/m$ expansion.
On the other hand, even though the HF and FW propagators go on-shell for the 
same four-momentum, they will differ off-shell.
Also this difference is captured by the field redefinition.
In order to do the field redefinition it is necessary to 
know the lower components of the original, relativistic Lagrangian.

\begin{acknowledgments}
We appreciate discussions with Daniel Phillips and Harold Fearing.
This work was supported in part by the NSF grant No.\ PHY-0457014.
\end{acknowledgments}

\appendix*
\section{Integrating out small components}
We repeat here the arguments given in Ref.~\cite{Mannel} (see also \cite{Das}) 
showing that the integration over the lower components of the HF Lagrangian 
gives a constant.
The discussion automatically carries over to the FW case. 
The path integral we need to perform for HF is
\begin{eqnarray}
  \mathcal{I}_{\rm HF} & = & \int[d{h'}^\dagger][dh']
  \exp\left[i\int d^4x {h'}^\dagger(2m+i\mathcal{D}_0-i\epsilon)h'\right] 
  \nonumber \\
  & = & {\rm Det}(2m+i\partial_0+i\Gamma_0-i\epsilon) 
  = {\rm Det}(2m+i\partial_0)
  {\rm Det}\left(1+\frac{1}{2m+i\partial_0-i\epsilon}\,i\Gamma_0\right)
  \nonumber \\
  & = & {\rm Det}(2m+i\partial_0)\exp\left[\frac{1}{2}{\rm Tr}\ln\left(
    1+\frac{1}{2m+i\partial_0-i\epsilon}\,i\Gamma_0\right)\right]={\rm Const.}
\end{eqnarray}
As argued in Ref.~\cite{Mannel}, the propagator in the logarithm is
always propagating in the backward time direction only.
Thus the trace can not be closed, it has to vanish, and the determinant is a 
constant as indicated.

To $\mathcal{O}(1/m^2)$ in FW we have
\begin{eqnarray}
  \mathcal{I}_{\rm FW} & = & \int[dn^\dagger][dn]
  \exp\left[i\int d^4xn^\dagger\left(2m+i\mathcal{D}_0+A
    +\frac{1}{2m}(-i\vec\sigma\cdot\vec\mathcal{D}+C)
    (-i\vec\sigma\cdot\vec\mathcal{D}+B) \right.\right. \nonumber \\
    & + & \frac{1}{4m^2}(-i\vec\sigma\cdot\vec\mathcal{D}+C)(i\mathcal{D}_0+A)
    (-i\vec\sigma\cdot\vec\mathcal{D}+B) \nonumber \\
    & - & \left.\left.
    \frac{1}{8m^2}\{(-i\vec\sigma\cdot\vec\mathcal{D}+C)
    (-i\vec\sigma\cdot\vec\mathcal{D}+B),(i\mathcal{D}_0+D)\}-i\epsilon
    \right)n\right].
\end{eqnarray}
The resulting determinant can, just as in the HF case, be factorized into the 
free propagator and interaction terms:
\begin{eqnarray}
  \mathcal{I}_{\rm FW} & = & 
  {\rm Det}\left(2m+i\partial_0-\frac{\nabla^2}{2m}-\frac{\nabla^4}{8m^3}+
  \ldots\right) \nonumber \\
  & \times & {\rm Det}\left[1+\frac{1}{2m+i\partial_0-\frac{\nabla^2}{2m}-
      \frac{\nabla^4}{8m^3}+\ldots-i\epsilon}({\it interaction\ terms})\right]
  \nonumber \\
  & = & {\rm Det}\left(2m+i\partial_0-\frac{\nabla^2}{2m}-
  \frac{\nabla^4}{8m^3}+\ldots\right) \nonumber \\
  & \times & \exp\left[\frac{1}{2}{\rm Tr}\ln\left(
    1+\frac{1}{2m+i\partial_0-\frac{\nabla^2}{2m}-
      \frac{\nabla^4}{8m^3}+\ldots-i\epsilon}({\it interaction\ terms})
    \right)\right] = {\rm Const.}, \nonumber \\
\end{eqnarray}
where in the last step we again use that the propagator propagates in the
backward time direction only, so that no closed loops are possible.

\bibliographystyle{apsrev}

\end{document}